\documentclass[5p,times]{elsarticle}

\usepackage{breakcites}
\usepackage{lineno}
\usepackage{graphicx}
\usepackage{subfigure}
\usepackage{amsmath}
\usepackage{algorithm}
\usepackage{algorithmic}
\usepackage{multirow}
\usepackage{setspace}
\usepackage{stfloats}

\begin{document}

\sloppy

\begin{frontmatter}

\title{Amplification Trojan Network: Attack Deep Neural Networks by Amplifying Their Inherent Weakness}


\author[tsinghua_address]{Zhanhao~Hu}

\author[tsinghua_address]{Jun~Zhu}

\author[tsinghua_address]{Bo~Zhang}

\author[tsinghua_address,IDG_address,CIBR_address]{Xiaolin~Hu\corref{mycorrespondingauthor}}
\cortext[mycorrespondingauthor]{Corresponding author}
\ead{xlhu@tsinghua.edu.cn}

\address[tsinghua_address]{Department of Computer Science and Technology, Institute for Artificial Intelligence, State Key Laboratory of Intelligent Technology and Systems, THBI, BNRist, Tsinghua University, Beijing 100084, China.}
\address[IDG_address]{IDG/McGovern Institute for Brain Research, Tsinghua University, Beijing 100084, China.}
\address[CIBR_address]{Chinese Institute for Brain Research, Beijing, China.}

\begin{abstract}
Recent works found that deep neural networks (DNNs) can be fooled by adversarial examples, which are crafted by adding adversarial noise on clean inputs. The accuracy of DNNs on adversarial examples will decrease as the magnitude of the adversarial noise increase. In this study, we show that DNNs can be also fooled when the noise is very small under certain circumstances. This new type of attack is called Amplification Trojan Attack (ATAttack). Specifically, we use a trojan network to transform the inputs before sending them to the target DNN. This trojan network serves as an amplifier to amplify the inherent weakness of the target DNN. The target DNN, which is infected by the trojan network, performs normally on clean data while being more vulnerable to adversarial examples. Since it only transforms the inputs, the trojan network can hide in DNN-based pipelines, e.g. by infecting the pre-processing procedure of the inputs before sending them to the DNNs. This new type of threat should be considered in developing safe DNNs.
\end{abstract}


\end{frontmatter}


\section{Introduction}

Deep neural networks (DNNs) have achieved good success in different fields. However, recent researches show that most DNNs are vulnerable to adversarial examples~\cite{goodfellow2014explaining,szegedy2013intriguing}. Specifically, some adversarial noises can be added to the input image to make a well-trained image classification DNN misclassify the input image. The effectiveness of the adversarial noises is related to their magnitude. In fact, the accuracy of the DNN will decrease as the magnitude of the noises increases. Many attack methods~\cite{szegedy2013intriguing,goodfellow2014explaining,kurakin2016adversarial,papernot2016the,nguyen2015deep,moosavi2016deepfool,carlini2017towards,dong2018boosting} are proposed to craft adversarial examples. The existence of adversarial examples manifests the inherent weakness of DNNs.

\begin{figure*}[t]
  \centering
  \subfigure[]{
  \includegraphics[width=0.8\linewidth]{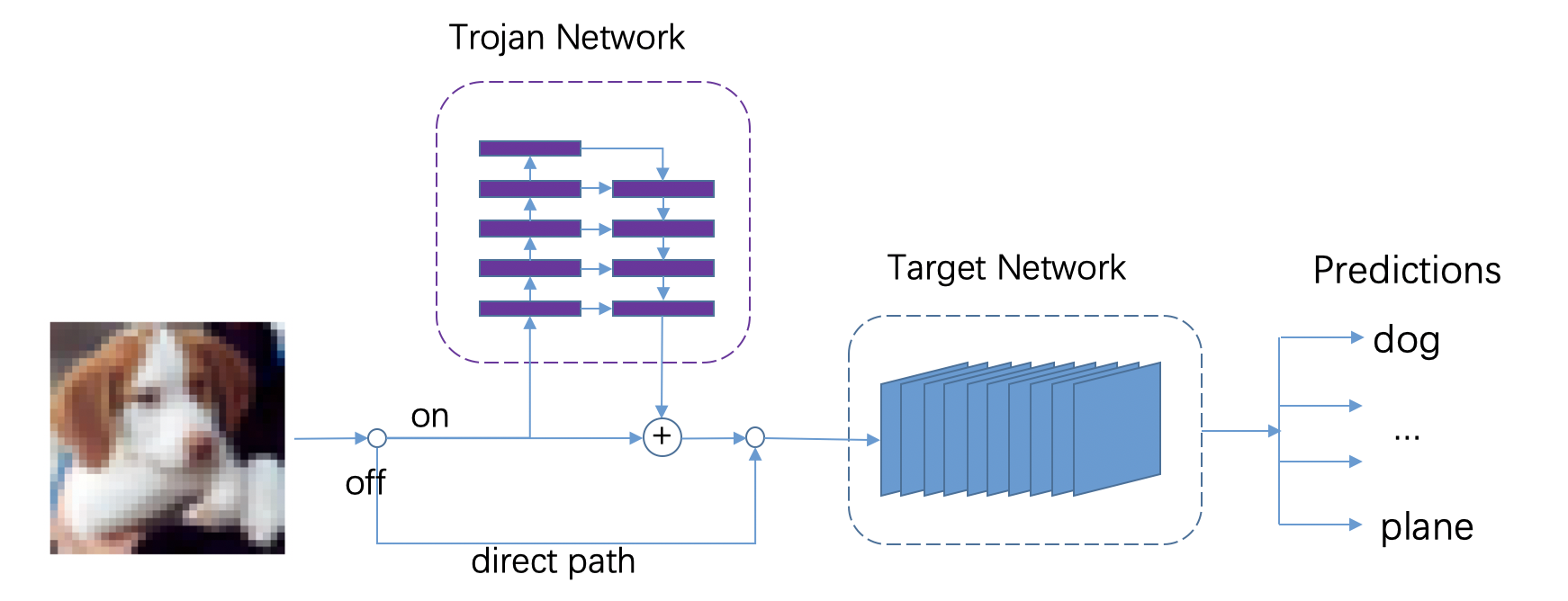}
  \label{fig:model}
  }
  \subfigure[]{
  \includegraphics[width=0.8\linewidth]{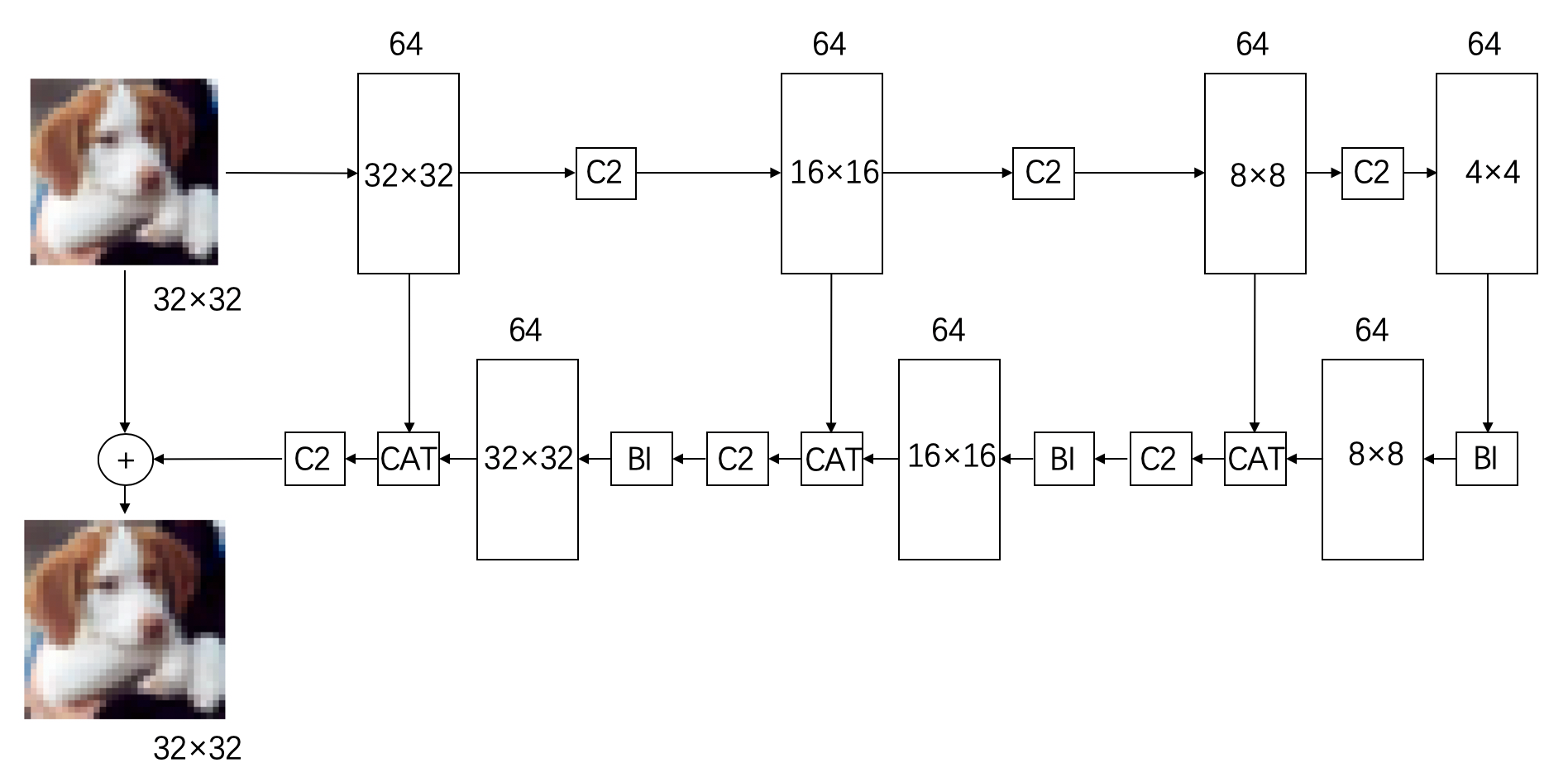}
  \label{fig:trojan}
  }
  \caption{The pipeline of ATAttack and the architecture of ATNet. \subref{fig:model} Pipeline of the attack. ATNet can be switched on or off. When it is switched off, the input is directly sent to the host network. When it is switched on, it transforms the input then send the result to the host network. \subref{fig:trojan} Architecture of ATNet. We denote \emph{C2} as two stacked convolutional layers, \emph{BI} as a bilinear interpolation upsampling layer, and \emph{CAT} as a concatenation operation. The numbers inside the boxes denote the spatial size of the feature maps and the numbers on top of the boxes denote the number of feature maps.}
\end{figure*}


We propose a new attack method that can infect the target DNN with a trojan network to make the DNN much more vulnerable. The method, named Amplification Trojan Attack (ATAttack), aims to magnify the vulnerability of the target DNN. Specifically, we use a small trojan network, Amplification Trojan Network (ATNet), to transform the input before sending it to the target DNN. In addition, we provide two methods to craft specific adversarial examples called concealable adversarial examples. The pipeline of our attack is shown in Fig.~\ref{fig:model}. The architecture of ATNet is shown in Fig.~\ref{fig:trojan}, which is inspired by Liao et al.~\cite{liao2018defense}. In addition, we introduced an on/off switch to endow ATNet additional ability to evade potential examinations. When ATNet is switched on, the target network becomes more vulnerable and can be easily attacked by the concealable adversarial examples, but still performs normally on the clean examples. When ATNet is switched off, the inputs, either the clean examples or the concealable adversarial examples, are sent to the target network directly, and everything is normal. Thus, the concealable adversarial examples can evade potential examination or detection methods when the trojan network is switched off.

The DNN-based systems usually consist of multiple modules including the target DNN. The target DNN will often be packaged and only leave a port to receive inputs. Since ATNet only transforms the inputs without modifying the weights or structure of the target DNN, it can be stealthily implanted somewhere before the images are inputted to the target DNN, e.g., image inputting module, image preprocessing module. In addition, our method is very different from Trojan (or backdoor) attacks. Trojan attacks usually either change the weights \cite{gu2017badnets,liu2017trojaning,shafahi2018poison,saha2020hidden} or change the structure of the target DNN~\cite{Tang2020An}. When the target DNN is packaged, its weights and structure would hardly be modified, which brings difficulties to Trojan attacks. For example, Tang et al.~\cite{Tang2020An} proposed to insert a small Trojan network into the target DNN and modify its penultimate layer. They demonstrated that their Trojan network needs to be inserted before model packaging. ATNet, in contrast, does not have this limitation. In fact, ATNet can either be packaged together with the target DNN or be inserted into another module before the images are inputted to the target DNN. As an example, a DNN-based automatic monitoring system first captures images by a digital camera and processes the raw signals by several modules (including an image sensor, an A/D converter, and an image processing engine \cite{Lukac2006}). The processed images are then inputted into a DNN-based system. In this case, one can either insert the ATNet into the DNN-based system or insert the ATNet into the image processing engine. That is to say, ATNet can be implant more flexibly.


By using this method, we break the typical DNNs widely used in the computer vision field, even when we applied one of the strongest defense methods, adversarial training~\cite{madry2017towards,wong2020fast}, to improve their robustness.


\section{Related works}

\subsection{Adversarial Attacks}

Various adversarial attacks~\cite{goodfellow2014explaining,szegedy2013intriguing,kurakin2016adversarial,papernot2016the,nguyen2015deep,moosavi2016deepfool,carlini2017towards,dong2018boosting} have been proposed to craft adversarial examples under different situations. Goodfellow et al.~\cite{goodfellow2014explaining} introduced the Fast Gradient Sign Method (FGSM). The perturbation is calculated by taking a single step gradient descent. It is a fast and straightforward way to craft adversarial examples. Kurakin et al.~\cite{kurakin2016adversarial} later proposed the Basic Iterative Method (BIM). It is an extension of the FGSM that takes more steps to get an adversarial example.


The FGSM and BIM are closely related to our proposed method. Thus we briefly describe them here.

\subsubsection{FGSM}
The FGSM~\cite{goodfellow2014explaining} takes one step by the sign of the gradient of the inputs. The perturbation $\delta$ is
given by
\begin{equation}
  \delta = \epsilon \mathrm{Sign}(\nabla_x\mathcal{J}(\theta, x, l)),
\end{equation}
where $\epsilon$ is a small value that restricts the $l_\infty$ norm of the perturbation. $\theta$ is the fixed parameters of the model. $x$ is the input and $l$ is the corresponding label. $\mathcal{J}$ is the cost function of the model. The adversarial example $x_\epsilon$ is then crafted by
\begin{equation}
  x_\epsilon = x + \delta.
\end{equation}

\subsubsection{BIM-K}
BIM-K~\cite{kurakin2016adversarial} takes $K$ steps to find an adversarial example. It iteratively updates the input as follows

\begin{equation}
  x_\epsilon^{(i+1)} = \mathrm{Clip}_{X, \epsilon}(x_\epsilon^{(i)} + \alpha \mathrm{Sign}(\nabla_{x_\epsilon^{(i)}}{\mathcal{J}(\theta, x_\epsilon^{(i)}, l)})),
\end{equation}
where $i\in[0,1,...,K-1]$ and $x_\epsilon^{(i)}$ is the adversarial example in the $i$th iteration. $\mathrm{Clip}_{X, \epsilon}(.)$ is a function that performs pixel-wised clipping
\begin{equation}
  \mathrm{Clip}_{X, \epsilon}(x) = \max(1, X+\epsilon, \min(0, X-\epsilon, x)),
\end{equation}
where $X$ is the original image and its pixels are normalized to $[0,1]$.



\subsection{Adversarial Training}

Adversarial training~\cite{szegedy2013intriguing,goodfellow2014explaining,madry2017towards,wong2020fast} is an effective defense method to protect a target model from adversarial attacks. The main idea is to augment the training set by incorporating adversarial examples generated by certain adversarial attack methods when training. It becomes a baseline defense method for evaluating new attack methods.

The object function for adversarial training is
\begin{equation}
  \tilde{\mathcal{J}}(\theta, x, l) = \beta\mathcal{J}(\theta, x, l) + (1-\beta)\mathcal{J}(\theta, x_a, l),
\end{equation}
where $x_a$ denotes the adversarial example that is generated from clean input $x$, and $\beta$ is a coefficient between $0$ and $1$.

\subsection{High-Level Representation Guided Denoiser}
Liao et al.~\cite{liao2018defense} proposed High-level Representation Guided Denoiser (HGD) to defend against adversarial attacks. They put a denoiser network before the target network and trained the denoiser network on both clean and adversarial examples while the target network remained fixed.


We use a similar model to the HGD. However, our model is trained to amplify the adversarial noises' influence on the target network instead of reducing it.

\section{Amplification Trojan Attack}

\subsection{Attack Pipeline}

The pipeline of our attack is shown in Fig. \ref{fig:model}. ATNet is a small network that transforms the inputs before sending them to the target network. ATNet can be switched on or off.  When ATNet is switched on, the inputs will be transformed before being sent to the target network. The target network will recognize the transformed clean examples as usual but misclassify the transformed adversarial examples even when the magnitude of the adversarial noise is tiny. When it is switched off, the inputs will be directly sent to the target network. If there is no way to switch off ATNet, when an inspector acquires the crafted adversarial examples and inputs them to the system, he/she will discover the abnormity of the system as the system outputs clearly wrong results. We provide an option of switching off ATNet (in practice, this should be the setting in most of the time), then the inspector cannot find any abnormity as the system outputs normal results.

The architecture of ATNet is shown in Fig.~\ref{fig:trojan}, which is inspired by Liao et al.~\cite{liao2018defense}. They proposed denoising U-net (DUNET) to remove the adversarial noise in the input images. Since we also manipulate the input image, but with a different purpose——amplifying the adversarial perturbation, we design a similar structure for ATNet. It is well-known that U-net structure is good at tasks in which input and output have the same dimension. This is the motivation for using this structure.

ATNet only transforms the inputs without modifying the weights or structures of the target DNN, which means that it is not coupled with the target DNN and can be flexibly implanted to a DNN-based system. Since the ATNet can be implanted somewhere before the images are inputted to the target DNN, the attacker can choose to implant the ATNet somewhere easy to access in the entire system.

\subsection{Requirements of the attack}

We denote the target network as $F(\cdot)$ which maps the input image $x$ to the logits $F(x)$ (the penultimate layer before the softmax function). We denote ATNet as $G(\cdot)$. Therefore, when ATNet is switched on, the final output becomes $F(G(x))$. We denote the clean example by $x_c$ and the trigger by $\delta$. The ground-truth label of the input is denoted by $l$, and the target label is denoted by $l^*$. There are five requirements to deploy a successful ATAttack:

\begin{eqnarray}
  \label{eq:basic}
  \arg \max F(x_c) &=& l \\
  \label{eq:identity}
  \arg \max F(G(x_c)) &=& l \\
  \label{eq:hiding}
  \arg \max F(x_c+\delta) &=& l \\
  \label{eq:backdoor}
  \arg \max F(G(x_c+\delta)) &=& l^* \\
  \label{eq:imperceptibility}
  ||\delta||_p &\le& \epsilon.
\end{eqnarray}
Eq.~(\ref{eq:basic}) denotes the target network's \emph{basic requirement}, which requires the target network to classify the clean example correctly. Eq.~(\ref{eq:identity}) denotes the \emph{identity requirement}, which means that ATNet preserve the target network's accuracy when the example is clean. Eq.~(\ref{eq:hiding}) denotes the \emph{concealment requirement}, which requires the adversarial examples not to influence the output of the target network when ATNet is switched off. Eq.~(\ref{eq:backdoor}) denotes the \emph{attack requirement}, which defines the target of the adversarial attack. Additionally, we have Eq.~(\ref{eq:imperceptibility}) as the \emph{imperceptibility requirement}, which restricts the $l_p$-norm of the perturbation to a constant $\epsilon$. In this work, we consider $l_\infty$-norm only. Since the target network is pre-trained and fixed, we only need to consider the latter four requirements (\ref{eq:identity})-(\ref{eq:imperceptibility}) when implementing ATAttack.

\subsection{Crafting Concealable Adversarial Examples}

We use two attack algorithms to craft \emph{concealable} adversarial examples based on FGSM~\cite{goodfellow2014explaining} and BIM-K~\cite{kurakin2016adversarial}. Compared to the original algorithms that only aim to satisfy the attack requirement, our concealable variants aim to satisfy both attack and concealment requirements. Therefore, the crafted adversarial examples will not influence the accuracy of the target network when ATNet is switched off, which can help both ATNet and the adversarial examples evade potential examination or detection methods. Since Eq.~(\ref{eq:identity}) does not involve $\delta$, we only need to consider three requirements: concealment requirement~(\ref{eq:hiding}), attack requirement~(\ref{eq:backdoor}) and imperceptibility requirement~(\ref{eq:imperceptibility}) when crafting concealable adversarial examples.

\subsubsection{Concealable FGSM}

A direct idea to satisfy the three requirements by one step is to plus an additional loss $L_h'$ that can help satisfy the concealment requirement onto the original loss of the FGSM. However, it cannot be directly applied here because its gradient will not work. For instance, a straightforward idea to define $L_h'$ is
\begin{equation}
  L_h' = ||F(x_c + \delta)-F(x_c)||_2^2.
  \label{eq:Lh_another}
\end{equation}
Here we consider what happens when FGSM is applied. Note that FGSM requires the gradient of the input. The input's gradient is the summation of the original gradient and the gradient to $L_h'$, where the gradient to $L_h'$ is
\begin{equation}
  \nabla_{\delta} L_h' = 2(F(x_c + \delta)-F(x_c))\nabla_{\delta}(F(x_c + \delta)).
\end{equation}
It is zero when $\delta=0$, which is the start point of the FGSM. Thus $L_h'$ does not guide the adversarial perturbation.

To overcome this problem, we propose Concealable FGSM (C-FGSM) to craft the adversarial examples. We use two losses for the attack requirement and concealment requirement and compute the gradients independently. Next, we compute the projection of the two gradients to compute the final direction. Finally, we take a small step in the final direction to satisfy the imperceptibility requirement. This process is detailed as follows.

We denote the concealment requirement loss as $L_h$ and the attack requirement loss as $L_b$:
\begin{eqnarray}
  &&L_h = \mathrm{CELoss}(F(x_c + \delta), l)
  \label{eq:Lh}
  \\
  &&L_b = \mathrm{CELoss}(F(G(x_c + \delta)), l^*),
  \label{eq:Lb}
\end{eqnarray}
where $\mathrm{CELoss}(.)$ denotes the Cross Entropy Loss. $l$ can be set to the predicted label of the target network when the ground-truth is unknown. Our goal is to minimize $L_b$ (\ref{eq:Lb}) and keep $L_h$ (\ref{eq:Lh}) unchanged. We denote the gradients of $L_h$ and $L_b$ with respect to $\delta$ as $g_h$ and $g_b$, respectively:
\begin{eqnarray}
  &&g_h = \nabla L_h
  \label{eq:gh}
  \\
  &&g_b = \nabla L_b.
  \label{eq:gb}
\end{eqnarray}
$L_h$ (\ref{eq:Lh}) does not change much along the direction which is perpendicular to $g_h$, while $L_b$ (\ref{eq:Lb}) decreases along the direction of $-g_b$. We denote $g_\parallel$ as the component of $g_b$ that is parallel to $g_h$ and denote $g_\perp$ as the component of $g_b$ that is perpendicular to $g_h$:
\begin{eqnarray}
  && g_\parallel = \frac{g_h g_b^T g_h}{||g_h||_2^2} \\
  && g_\perp = g_b - g_\parallel.
\end{eqnarray}
\begin{figure}[t]
  \centering
  \includegraphics[width=0.6\linewidth]{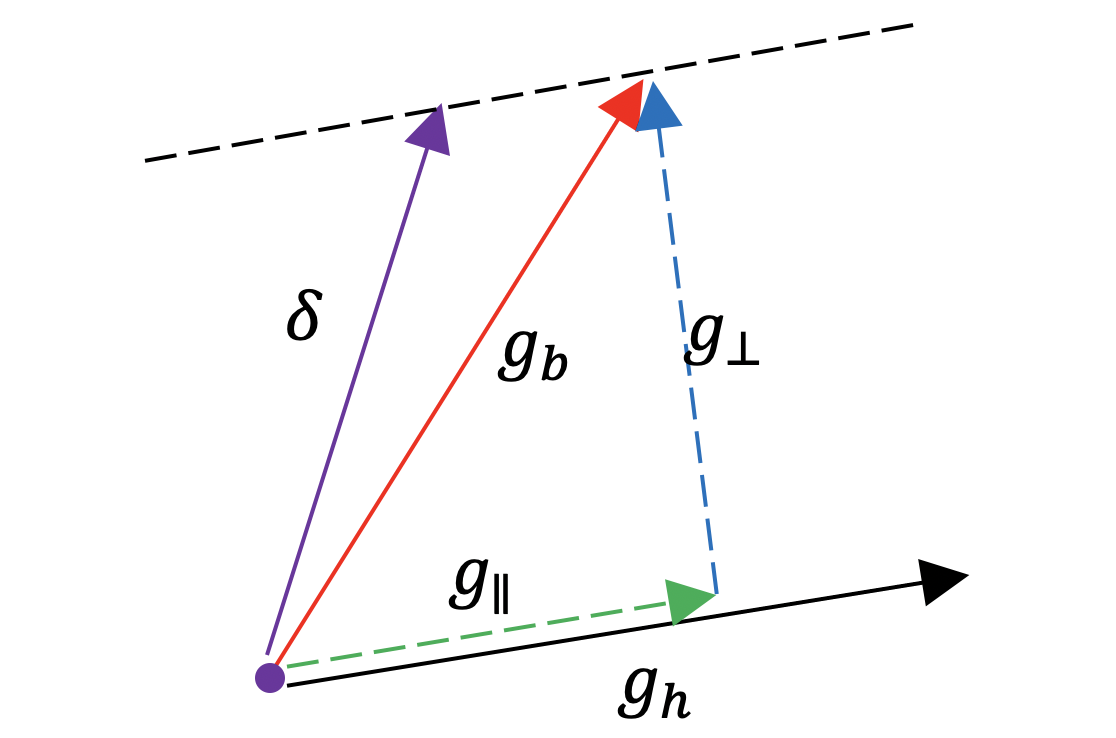}
  \caption{The relationship between the gradients. The black arrow represents $g_h$, the red arrow represents $g_b$, the green arrow represents $g_\parallel$ and the blue arrow represents $g_\perp$. $g_\parallel$ is the component of $g_b$ that is parallel to $g_h$ and $g_\perp$ is the component of $g_b$ that is perpendicular to $g_h$. $\delta$ is the summation of $(1-\lambda)g_\parallel$ and $g_\perp$, where $\lambda\in [0,1]$. The black dashed line is parallel to $g_h$.}
  \label{fig:g12}
\end{figure}
The adversarial perturbation $\delta$ is then computed by
\begin{equation}
  \label{eq:PFGSM}
  \delta = -\epsilon\mathrm{Sign}(g_\perp + (1 - \lambda) g_\parallel),
\end{equation}
where $\lambda$ is a coefficient, and $\epsilon$ is the upper bound of the $L_\infty$ norm of the perturbation. The corresponding adversarial example is $x_\epsilon = x_c + \delta$. When $\lambda=0$, the perturbation goes directly along the direction of $-g_b$ and $L_h$ is ignored. In this case, it is equivalent to the original FGSM. When $\lambda=1$, the perturbation goes along the direction that is perpendicular to $g_h$. The relationship between the gradients is illustrated in Fig.~\ref{fig:g12}. We found $\lambda=1$ was good for all of our experiments.

By replacing (\ref{eq:Lb}) with
\begin{equation}
  L_b = -\mathrm{CELoss}(F(G(x_c + \delta)), l),
  \label{eq:untargeted}
\end{equation}
we can use this method to perform untargeted attack.

\subsubsection{Concealable BIM-K}

Since BIM-K is a multi-step optimization method, $L_h'$ (\ref{eq:Lh_another}) is effective to satisfy the concealment requirement. Thus, we minimize $L_h'$  to satisfy the concealment requirement and minimize $L_b$ ((\ref{eq:Lb}) for targeted attack and (\ref{eq:untargeted}) for untargeted attack) to satisfy the attack requirement. We denote this variant as Concealable BIM-K (C-BIM-K). The adversarial perturbation $\delta$ is computed by
\begin{equation}
  \label{eq:step1_1}
  \delta = \mathop{\arg\min} \limits_{||\delta||_\infty < \epsilon} c_h \cdot L_h + L_b,
\end{equation}
where $c_h$ is a coefficient. We iteratively solve this problem by
\begin{equation}
  \label{eq:BIM}
  \begin{aligned}
    \delta^{(i+1)} = &\mathrm{Clip}_{x_c, \epsilon}(\delta^{(i)} - \alpha\mathrm{Sign}(\nabla_{x_c + \delta^{(i)}}{(c_h \cdot L_h + L_b)})),
  \end{aligned}
\end{equation}
where $i\in[0, 1, .., K-1]$ and $\alpha$ is the step size in each iteration. It loops for $K$ times in total. The final adversarial example is crafted by $$x_\epsilon = x_c + \delta^{(K)}$$

\subsection{Training ATNet}

When training ATNet, we only need to consider the identity requirement~(\ref{eq:identity}) and the attack requirement~(\ref{eq:backdoor}). We minimize identity Loss $L_i$ to satisfy the identity requirement and minimize $L_b$ (\ref{eq:Lb}) to satisfy the attack requirement, where $L_i$ is defined as
\begin{equation}
  L_i = ||F(G(x_c)) - F(x_c)||_2^2.
\end{equation}
The parameters of ATNet $G$ is updated to optimize
\begin{equation}
  \label{eq:step2}
    \min_G c_i \cdot L_i + L_b,
\end{equation}
where $c_i$ is a coefficient. Since $G$ and $\delta$ are coupled in loss $L_b$, we fix one and optimize another alternately during training. In each cycle, we first compute the concealable adversarial examples and use Stochastic Gradient Descent (SGD) to update the weights of ATNet. Our training algorithm is similar to the standard adversarial training algorithm~\cite{madry2017towards} but with a different goal. See Algorithm \ref{algorithm1}.

\begin{algorithm}
  \caption{Training algorithm}
  \label{algorithm1}
  \begin{algorithmic}[1]
    \REQUIRE
    The pre-trained target network $F$; Training set $data$; total iterations $n$.
    \ENSURE
    ATNet $G$ and adversarial examples $\delta$
    \STATE Randomly initialize $G$
    \STATE $iters=0$
    \WHILE {$iters < n$}
    \STATE Sample input $x$ and the corresponding ground-truth label $l$ from $data$
    \STATE Select a target label $l^*\neq l$
    \STATE Fix $G$, generate $\delta$ by using C-FGSM(\ref{eq:PFGSM}) or C-BIM-$K$(\ref{eq:step1_1}) and get $x_{\epsilon} = x+\delta$
    \STATE Fix $x_{\epsilon}$, train $G$ by using (\ref{eq:step2})
    \STATE $iters = iters + 1$
    \ENDWHILE
  \end{algorithmic}
\end{algorithm}

\section{Experiments}

In this section, we first demonstrate implementation details of the experiments. Then, we investigate how well ATAttack satisfies the requirements (\ref{eq:identity})-(\ref{eq:imperceptibility}) on MNIST and CIFAR10, respectively. Meanwhile, we also compare ATAttack with the traditional adversarial attacks in terms of the accuracy of the target DNN on the adversarial examples. Next, we show the transferability of the ATNet by attacking unseen target DNNs on CIFAR10. In addition, we demonstrate the results of ATAttack on ImageNet. After that, we study the parameter sensitivity of the concealable adversarial attack methods. Finally, we analyze the amplification ability of ATNet.

\subsection{Implementation Details}

\subsubsection{Datasets}
We trained and tested our model on the MNIST, CIFAR10 and ImageNet datasets.
\paragraph*{MNIST} MNIST~\cite{lecun1998gradient} is a dataset of hand-writing numbers with $10$ categories. It has $60000$ training images and $10000$ test images. Each image is a gray scale image with $28*28$ pixels.

\paragraph*{CIFAR10} CIFAR10~\cite{krizhevsky2009learning} has $10$ categories. It has $50000$ training images and $10000$ test images. Each image is an RGB image with $32*32$ pixels.

\paragraph*{ImageNet} ImageNet is a large dataset, and the most commonly used sub-dataset is ISLVRC 2012~\cite{russakovsky2015imagenet}. In what follows, the ImageNet dataset refers to this sub-dataset. It has 1000 categories, and its training set contains over one million images. We constructed a small subset to train ATNet as follows. We randomly selected five images in the training set of ImageNet for each category, which constituted our training set. In the same way, we randomly selected five images in the validation set of ImageNet for each category to constitute our test set.

\subsubsection{Models} The target networks and the trojan networks were different for different datasets.

\begin{table}[b]
  \scriptsize
  \centering
  \caption{The structure of the CNN-small trained on the MNIST dataset}
  \begin{tabular*}{1.0\linewidth}{@{\extracolsep{\fill}}lll}
    \hline
    & \multicolumn{1}{l}{MNIST Model} \\
    \hline
    Layer Type & Kernel & Units \\
    \hline
    ReLU Convolution & $3\times3$ & $32$ \\
    ReLU Convolution & $3\times3$ & $32$ \\
    Max Pooling & $2\times2$ & / \\
    ReLU Convolution & $3\times3$ & $64$ \\
    ReLU Convolution & $3\times3$ & $64$ \\
    Max Pooling & $2\times2$ & / \\
    ReLU FC & / & $200$ \\
    ReLU FC & / & $200$ \\
    Softmax & / & $10$ \\
    \hline
  \end{tabular*}
  \label{table:structure}
\end{table}

\paragraph*{MNIST} We first trained a small CNN (CNN-small) with four convolutional layers on MNIST. The structure of the network is described in Table~\ref{table:structure}. The accuracy on the test set was $99.49\%$. We used a smaller network than ATNet, denoted by ATNet-small, as the trojan network. It was obtained by halving the number of channels in every layer of ATNet.

\paragraph*{CIFAR10} We trained and fixed the weights of Resnet18~\cite{he2016deep} on CIFAR10 and obtained $95.00\%$ accuracy on the test set. For the transfer study, we also trained Alexnet~\cite{krizhevsky2017imagenet}, VGG9~\cite{simonyan2014very} and CNN-small on CIFAR10 as the target networks. We reduced the kernel size of the first convolutional layer in Resnet, Alexnet and VGG9 to fit the input size of the images in the CIFAR10. Moreover, we also used adversarial training to train the Resnet18. We used ATNet as the trojan network.

\paragraph*{ImageNet} We used a pretrained Resnet50~\cite{he2016deep} as the target network on ImageNet. The pretrained model had been trained by adversarial training. We used ATNet as the trojan network.

\begin{table*}
  \scriptsize
  \centering
  \caption{The results of the target network CNN-small on MNIST with/without different trojan networks}
\begin{tabular*}{1.0\linewidth}{@{\extracolsep{\fill}}llllll}
\hline
\multirow{2}{*}{The target network}                & Clean  & FGSM  & BIM10UT   & \multicolumn{2}{c}{BIM10RT}   \\
                                                   & acc/\% & acc/\% & acc/\%    & acc/\%        & suc/\%        \\ \hline
CNN-small (No Trojan)                                          & 99.49  & 95.80  & 93.28     & 97.98         & 1.23          \\ \hline
\hline
\multirow{2}{*}{The trojan networks (switched on)} & Clean  & C-FGSM & C-BIM10UT & \multicolumn{2}{c}{C-BIM10RT} \\
                                                   & acc/\% & acc/\% & acc/\%    & acc/\%        & suc/\%        \\ \hline
ATNet-small-C-FGSM                                 & 98.77  & 6.96   & 9.10      & 1.43          & 84.99         \\
ATNet-small-C-BIM10UT                              & 99.43  & 11.64  & 6.71      & 10.50         & 51.50         \\
ATNet-small-C-BIM10RT                              & 99.02  & 9.76   & 0.77      & 2.79          & 42.57         \\ \hline
The trojan networks (switched off)                 & \multicolumn{5}{l}{}                                        \\ \hline
ATNet-small-C-FGSM                                 & 99.49  & 99.30  & 99.46     & 99.47         & 0.07          \\
ATNet-small-C-BIM10UT                              & 99.49  & 98.56  & 99.49     & 99.50         & 0.03          \\
ATNet-small-C-BIM10RT                              & 99.49  & 99.04  & 99.50     & 99.49         & 0.04          \\ \hline
\end{tabular*}
  \label{table:MNIST_trojan}
\end{table*}

\begin{table*}
  \scriptsize
  \centering
  \caption{The results of the target network Resnet18 on CIFAR10 with/without different trojan networks}
\begin{tabular*}{1.0\linewidth}{@{\extracolsep{\fill}}llllll}
\hline
\multirow{2}{*}{The target network}                & Clean  & FGSM   & BIM10UT   & \multicolumn{2}{c}{BIM10RT}   \\
                                                   & acc/\% & acc/\% & acc/\%    & acc/\%         & suc/\%        \\ \hline
Resnet18(No Trojan)                                         & 95.00  & 66.86  & 49.89     & 72.96          & 17.94         \\ \hline
\hline
\multirow{2}{*}{The trojan networks (switched on)} & Clean  & C-FGSM & C-BIM10UT & \multicolumn{2}{c}{C-BIM10RT} \\
                                                   & acc/\% & acc/\% & acc/\%    & acc/\%         & suc/\%        \\ \hline
ATNet-C-FGSM                                       & 94.06  & 9.56   & 8.96      & 8.14           & 78.63         \\
ATNet-C-BIM10UT                                    & 94.46  & 18.15  & 5.75      & 7.21           & 84.78         \\
ATNet-C-BIM10RT                                    & 94.28  & 44.57  & 3.31      & 7.10           & 86.52         \\ \hline
The trojan networks (switched off)                 & \multicolumn{5}{l}{}                                        \\ \hline
ATNet-C-FGSM                                       & 95.00  & 94.28  & 94.57     & 95.02          & 0.45          \\
ATNet-C-BIM10UT                                    & 95.00  & 93.50  & 94.70     & 95.19          & 0.48          \\
ATNet-C-BIM10RT                                    & 95.00  & 95.23  & 94.89     & 95.12          & 0.42          \\ \hline
\end{tabular*}
  \label{table:CIFAR10_trojan}
\end{table*}

\subsubsection{Training settings} We fixed all pre-trained models in later experiment and trained the trojan networks with Algorithm \ref{algorithm1}. The trojan networks were trained for 20 epochs by SGD with learning rate decaying from $0.001$ to $0.0001$.

\subsubsection{Attack settings} We tested the accuracy of the target networks on adversarial examples crafted by FGSM and BIM-K. We used FGSM for untargeted attacks only and used BIM-K for both untargeted and targeted attacks. We used C-FGSM and C-BIM-K to craft concealable adversarial examples when the trojan network was employed. Note that FGSM and C-FGSM are both one-step gradient-based attack algorithms and can be compared fairly. When performing BIM-K and C-BIM-K, we ran ten steps in most of our experiments. In addition, we restricted the adversarial noise by $||\delta||_\infty\leq0.05$ for MNIST, $||\delta||_\infty\leq0.004$ for CIFAR10 and $||\delta||_\infty\leq0.01$ for ImageNet by default.

\subsubsection{Other hyper-parameters} We set $\lambda = 1$ in Eq.~(\ref{eq:PFGSM}) and $c_h = 0$ in Eq.~(\ref{eq:BIM}) by default. More analysis will be found in Section~\ref{section:parameter}.  We set different values for $c_i$ in Eq.~(\ref{eq:step2}) when implementing different attack algorithms. The details will be found in Section \ref{section:main}.

\subsection{Main Results on MNIST and CIFAR10}
\label{section:main}

We tested the accuracy of the target networks on MNIST and CIFAR10. Meanwhile, we computed the accuracy of the target networks on the adversarial examples which original FGSM and BIM crafted on the test set of each dataset. We only used FGSM to implement an untargeted attack (denoted by FGSM), while we used BIM-K to implement both targeted (denoted by BIM10RT since we set $K=10$) and untargeted attacks (denoted by BIM10UT). We randomly selected a target label from the nine labels other than the ground truth label for the targeted attack. The results on MNIST and CIFAR10 are presented in the top rows of Table~\ref{table:MNIST_trojan} and Table~\ref{table:CIFAR10_trojan} respectively. In the tables, the \emph{acc} denotes the classification accuracy of the host network, and the \emph{suc} denotes the success rate of the network for classifying the examples as the selected target labels. The success rate is only meaningful for the targeted attack. The results showed that the original attack algorithms had limited ability to attack those target networks when the magnitude of the adversarial perturbation was small.

We then employed the ATAttack on MNIST and CIFAR10, respectively. On MNIST, the target model was CNN-small, and the trojan network was ATNet-small. On CIFAR10, the target model was Resnet18, and the trojan network was ATNet. We fixed the target network and trained the trojan network according to Algorithm~\ref{algorithm1}. In the algorithm, the adversarial examples were generated by either C-FGSM or C-BIM-K. We set $\lambda=1$ for C-FGSM, and set $c_h=0$ and $K=10$ for C-BIM-K during training and evaluation. We only used C-FGSM to apply untargeted attacks (denoted by C-FGSM), and used C-BIM-K to apply both targeted (denoted by C-BIM10UT) and untargeted attacks (denoted by C-BIM10RT). During training, in order to balance the requirements (\ref{eq:identity}) and (\ref{eq:backdoor}), we varied the value of $c_i$ in (\ref{eq:step2}) between $0$ and $1000$, and found a good enough setting: $c_i=100$, $500$ and $150$ for C-FGSM, C-BIM10UT and C-BIM10RT, respectively.

After training, we evaluated how well the requirements (\ref{eq:identity})-(\ref{eq:imperceptibility}) are satisfied by ATAttack. Specifically, the identity requirement \eqref{eq:identity} can be measured by the accuracy of the target network on the clean dataset when the trojan network is switched on (see \emph{clean} column in Table~\ref{table:MNIST_trojan} and Table~\ref{table:CIFAR10_trojan}. The accuracy is supposed to be close to the original clean accuracy of the target network. The concealment requirement \eqref{eq:hiding} and attack requirement \eqref{eq:backdoor} can be evaluated by the concealable adversarial examples crafted by the C-FGSM, C-BIM10UT or C-BIM10RT (see the corresponding column named by the attack algorithm in Table~\ref{table:MNIST_trojan} and Table~\ref{table:CIFAR10_trojan}. For concealment requirement \eqref{eq:hiding}, we calculated the accuracy of the target network on the concealable adversarial examples when the trojan network was switched off. This accuracy is also supposed to be close to the original clean accuracy. For attack requirement \eqref{eq:backdoor}, we calculated the accuracy of the target network on the concealable adversarial examples when the trojan network was switched on. Unlike the previous cases, this accuracy is supposed to be as low as possible. Moreover, we also calculated the success rate for C-BIM10RT. These results on MNIST and CIFAR10 are presented in Table~\ref{table:MNIST_trojan} and Table~\ref{table:CIFAR10_trojan} respectively. In the Tables, we name the trojan networks by the corresponding attack methods that generate adversarial examples during training. For instance, \emph{ATNet-small-C-FGSM} indicates that C-FGSM was employed to generate adversarial examples when we use Algorithm~\ref{algorithm1} to train the trojan network ATNet-small. We showed the results when the trojan network was switched on and off.

\begin{table}[t]
  \scriptsize
\centering
\caption{The results of the target network CNN-small on CIFAR10 with/without the trojan network}
\begin{tabular*}{1.0\linewidth}{@{\extracolsep{\fill}}lllll}
\hline
\multirow{2}{*}{The target network} & \multicolumn{2}{c}{DeepFool} & \multicolumn{2}{c}{C\&W} \\
                         & acc/\%      & mean $l_2$     & acc/\%    & mean $l_2$   \\ \hline
CNN-small (No Trojan)               & 0.50        & 1.990          & 0.00      & 1.549        \\ \hline
\hline
ATNet-small-C-FGSM      & 1.30        & 0.0655         & 0.00      & 0.0138       \\
ATNet-small-C-BIM10UT    & 1.10        & 0.0250         & 0.00      & 0.0429       \\
ATNet-small-C-BIM10RT    & 7.40        & 0.00930        & 0.00      & 0.0380       \\ \hline
\end{tabular*}
\label{table:MNIST_l2}
\end{table}

\begin{table}[t]
  \scriptsize
\centering
\caption{The results of the target network Resnet18 on CIFAR10 with/without the trojan network}
\begin{tabular*}{1.0\linewidth}{@{\extracolsep{\fill}}lllll}
\hline
\multirow{2}{*}{The target network} & \multicolumn{2}{c}{DeepFool} & \multicolumn{2}{c}{C\&W} \\
                         & acc/\%      & mean $l_2$     & acc/\%    & mean $l_2$   \\ \hline
Resnet18 (No Trojan)                & 4.4         & 0.660          & 0.00      & 0.184        \\ \hline
\hline
ATNet-C-FGSM       & 3.7         & 0.198          & 0.00      & 0.0550       \\
ATNet-C-BIM10UT    & 3.7         & 0.190          & 0.00      & 0.0497       \\
ATNet-C-BIM10RT    & 4.2         & 0.193          & 0.00      & 0.0567      \\ \hline
\end{tabular*}
\label{table:CIFAR_l2}
\end{table}

\begin{table*}[b]
  \scriptsize
\centering
\caption{The results of the target network Resnet18AT on CIFAR10 with/without the trojan network}
\begin{tabular*}{1.0\linewidth}{@{\extracolsep{\fill}}lllllll}
\hline
\multicolumn{2}{l}{\multirow{2}{*}{The target network}} & Clean  & FGSM   & BIM10UT   & \multicolumn{2}{c}{BIM10RT}   \\
\multicolumn{2}{l}{}                       & acc/\% & acc/\% & acc/\%    & acc/\%        & suc/\%        \\ \hline
\multicolumn{2}{l}{Resnet18AT(No Trojan)}                          & 90.69  & 86.26  & 86.17     & 89.63         & 1.94          \\ \hline
\hline
\multicolumn{2}{l}{\multirow{2}{*}{The trojan network}} & Clean  & C-FGSM & C-BIM10UT & \multicolumn{2}{c}{C-BIM10RT} \\
\multicolumn{2}{l}{}                       & acc/\% & acc/\% & acc/\%    & acc/\%        & suc/\%        \\ \hline
\multirow{2}{*}{ATNet-C-BIM10RT-AT}        & on         & 89.71  & 34.33  & 7.12      & 10.82         & 83.19         \\
                                           & off        & 90.69  & 90.58  & 90.26     & 90.71         & 1.07         \\
\hline
\end{tabular*}
\label{tabal:CIFAR10_advtrain}
\end{table*}

The results indicated that the identity requirement was almost satisfied by all the trojan networks because the clean accuracy of the target network hardly changed when the trojan networks were switched on. Meanwhile, the concealment requirement was also almost satisfied because the accuracies of the target network on the concealable adversarial examples hardly changed when the trojan networks were switched off. Moreover, when the trojan networks were switched on, the accuracies of the target network on the concealable adversarial examples were much lower than those when attacked by similar attacks. It indicated that the attack requirement was almost satisfied. As an example, in Table~\ref{table:MNIST_trojan}, the accuracy of the CNN-small on MNIST was $93.28\%$ when attacked by the BIM10UT algorithm. However, when it was infected by ATNet-small-C-BIM10RT, the accuracy was $0.77\%$ on the adversarial examples that were crafted by C-BIM10UT. In addition, the success rates of the target attack method increased from $1.23\%$ to $84.99\%$, $51.50\%$ and $42.57\%$ when CNN-small was infected by ATNet-small-C-FGSM, ATNet-small-C-BIM10UT and ATNet-small-C-BIM10RT, respectively. These results showed that ATAttack successfully broke the target networks.

ATNet can also amplify adversarial noises created by other adversarial attacks. See Table~\ref{table:MNIST_l2} and Table~\ref{table:CIFAR_l2} for the result of DeepFool~\cite{moosavi2016deepfool} and C\&W~\cite{carlini2017towards} attacks on MNIST and CIFAR10, respectively. Unlike FGSM and BIM-K, DeepFool and C\&W attacks both give priority to the attack success and loosely restrict the $l_2$ norm of the adversarial noises. Therefore, they cannot be compared directly with ATAttack that strictly restricts the $l_\infty$ norm. Instead, we compared the methods that combining ATNet and DeepFool/C\&W with the original adversarial attacks (DeepFool/C\&W without ATNet), by measuring the mean $l_2$ norm of the adversarial noises. From the table, the adversarial attacks need much smaller noises to attack the target networks with ATNet. For example, in Table~\ref{table:CIFAR_l2}, the mean $l_2$ norm of the adversarial noises created by C\&W attack on the target network without Trojan was $0.184$, while it decreased to $0.0497$ when the target network was implanted with ATNet-C-BIM10UT. Meanwhile, the accuracy of the target network on the adversarial examples were both $0.00\%$, which indicates that all the adversarial examples successfully misled the target network. Besides, the accuracies on the adversarial examples created by DeepFool slightly increased when the target network was implanted with ATNets. We attributed this to the destruction of the linear assumption of the DeepFool method, because we found the loss surface was more rugged with ATNet, which will be discussed in Section~\ref{sec:amplification}.

Moreover, we employed ATAttack to attack more robust models. As mentioned before, adversarial training is an effective method to defend against adversarial attacks. We were interested to know whether ATAttack could effectively break adversarially trained DNNs. First, we chose Resnet18 as the target network and used FGSM to adversarially train the network on CIFAR10 (following Wong et al.~\cite{wong2020fast}). We denoted the obtained network with Resnet18AT. Second, we used C-BIM10RT to generate adversarial examples and trained a new trojan network, ATNet-BIM10RT-AT. The evaluation results are presented in Table~\ref{tabal:CIFAR10_advtrain}.

As shown in the table, the network trained by adversarial training (Resnet18AT in Table~\ref{tabal:CIFAR10_advtrain}) was truly more robust than the network trained by the standard method (Resnet18 in Table~\ref{table:CIFAR10_trojan}). However, the Resnet18AT's accuracy on the adversarial examples crafted by the 10-iteration untargeted attack method dropped from $86.17\%$ (BIM10UT) to $7.12\%$ (C-BIM10UT). It indicated that ATAttack also broke the robust network Resnet18AT.

\begin{figure*}
  \centering
  \subfigure[]{
  \includegraphics[width=0.45\linewidth]{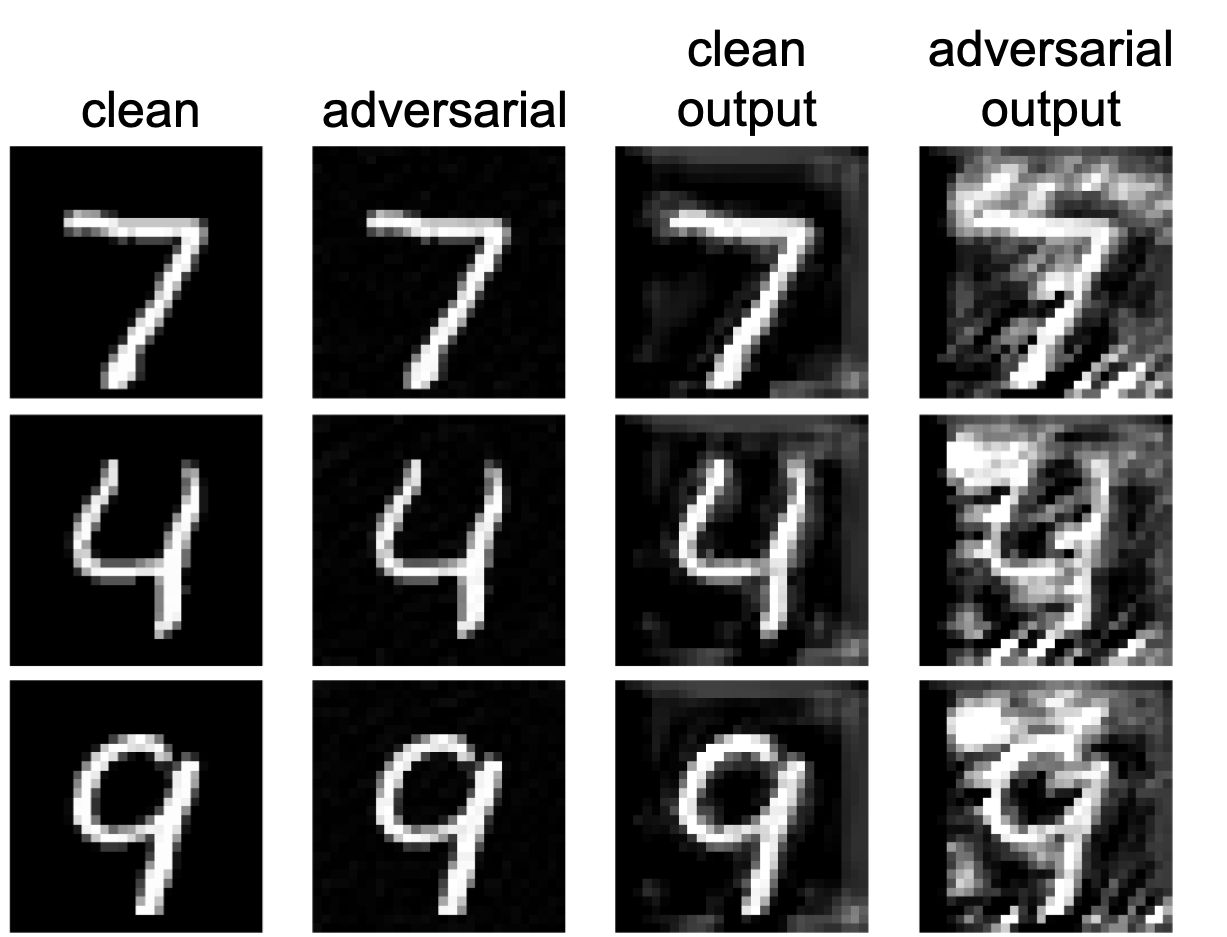}
  \label{fig:MNIST}
  }
  \subfigure[]{
  \includegraphics[width=0.45\linewidth]{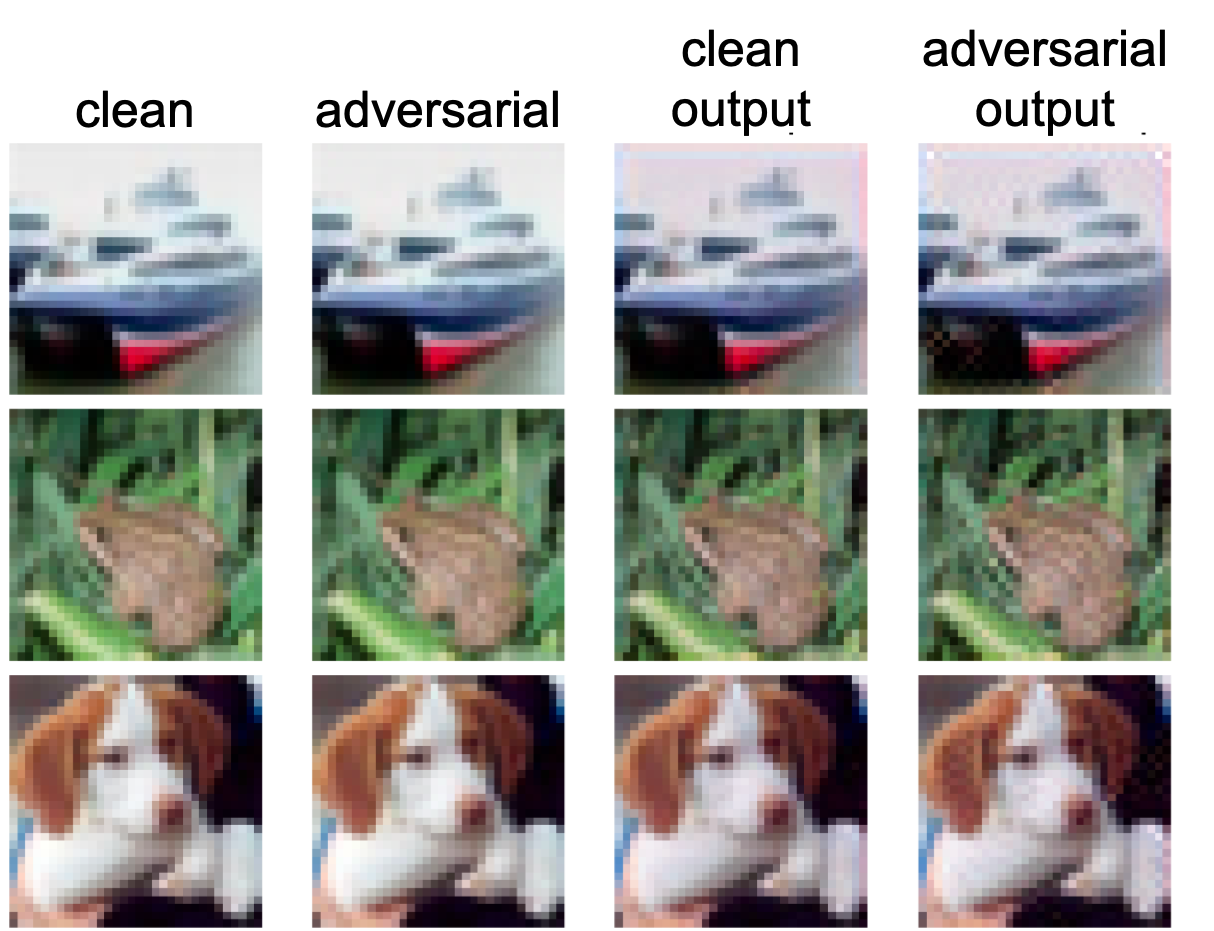}
  \label{fig:CIFAR10}
  }
  \caption{Visualization of the clean input, adversarial input and the corresponding output of the trojan networks. \subref{fig:MNIST} Visualization results on MNIST. \subref{fig:CIFAR10} Visualization results on CIFAR10. In each subfigure, the first column shows original clean examples, the second column shows the adversarial examples crafted by our proposed method, and the last two columns show the outputs of the trojan networks (switched on) for the inputs being the clean and adversarial examples respectively.}
  \label{fig:visual}
\end{figure*}

To give an intuitive vision to the transformation of the inputs by the trojan network, we visualized the clean and adversarial examples in Fig.~\ref{fig:visual}. The visualization results on the MNIST dataset are shown in Fig. \ref{fig:MNIST} and the CIFAR10 are shown in Fig. \ref{fig:CIFAR10}. The target network recognized all the adversarial examples in Fig. \ref{fig:MNIST} as the number $3$ when the trojan networks were switched on, while the adversarial examples shown in Fig. \ref{fig:CIFAR10} were all classified to \emph{cat} by the target network. These adversarial examples were all classified correctly when the trojan networks are switched off. The last two columns in Fig. \ref{fig:MNIST} and Fig. \ref{fig:CIFAR10} show that when the input was a clean image, the output of the trojan network was also a clean image; however, when the input was an adversarial image (though visually quite similar to the clean image), the output of the trojan network contained more noise. These results partially demonstrated the amplification effect of the trojan network. ATNet selectivity amplifies the potential noises in the input images. It hardly changes the input images when the images are legitimate examples, but amplifies the noise in the adversarial examples.

\begin{table*}
  \scriptsize
\centering
\caption{Transfer study of the trojan networks that are originally trained to attack Resnet18 on CIFAR10}
\begin{tabular*}{1.0\linewidth}{@{\extracolsep{\fill}}lllllll}
\hline
\multicolumn{2}{l}{\multirow{2}{*}{Target network}}               & Clean  & FGSM   & BIM10UT   & \multicolumn{2}{c}{BIM10RT}   \\
\multicolumn{2}{l}{}            & acc/\% & acc/\% & acc/\%    & acc/\%        & suc/\%        \\ \hline
\multicolumn{2}{l}{CNN-small}                                     & 91.02  & 36.63  & 26.36     & 62.66         & 32.32         \\
\multicolumn{2}{l}{Alexnet}                                       & 90.56  & 73.74  & 71.91     & 85.98         & 5.58          \\
\multicolumn{2}{l}{VGG9}                                          & 92.43  & 67.12  & 62.37     & 83.95         & 9.50          \\ \hline
\hline
\multirow{2}{*}{Target network} & \multirow{2}{*}{Trojan network} & Clean  & C-FGSM & C-BIM10UT & \multicolumn{2}{c}{C-BIM10RT} \\
                                &                                 & acc/\% & acc/\% & acc/\%    & acc/\%        & suc/\%        \\ \hline
\multirow{3}{*}{CNN-small}      & ATNet-C-FGSM                    & 89.54  & 8.47   & 26.64     & 3.39          & 88.28         \\
                                & ATNet-C-BIM10UT                 & 90.07  & 13.57  & 26.96     & 1.11          & 96.13         \\
                                & ATNet-C-BIM10RT                 & 90.06  & 22.73  & 22.71     & 0.89          & 97.07         \\
\hline
\multirow{3}{*}{Alexnet}        & ATNet-C-FGSM                    & 89.69  & 22.81  & 55.66     & 14.24         & 65.19         \\
                                & ATNet-C-BIM10UT                 & 90.18  & 46.26  & 61.16     & 19.80         & 61.38         \\
                                & ATNet-C-BIM10RT                 & 89.98  & 44.76  & 56.73     & 28.83         & 54.99         \\
\hline
\multirow{3}{*}{VGG9}           & ATNet-C-FGSM                    & 91.78  & 17.94  & 56.83     & 10.87         & 72.47         \\
                                & ATNet-C-BIM10UT                 & 91.91  & 37.63  & 57.17     & 9.92          & 77.93         \\
                                & ATNet-C-BIM10RT                 & 91.70  & 38.78  & 49.40     & 7.70          & 84.23         \\ \hline
\end{tabular*}
\label{tab:transfer}
\end{table*}

\subsection{Transferability of the Trojan Network}
\label{section:transfer}
We have shown that the trojan network can amplify the inherent weakness of the target network after training by Algorithm~\ref{algorithm1}. Since the trojan network was trained with a specific target network, there comes a question: can the trojan network that is trained to break one specific target network break other target networks?

We used Resnet18 as the target network to train the trojan networks ATNet on CIFAR10. The trojan network that used an attack algorithm to generate adversarial examples during training was named by the corresponding attack algorithm. For instance, the trojan network trained on adversarial examples generated by the C-FGSM was named \emph{ATNet-C-FGSM}. Since there were three attack algorithms, there were three trojan networks, ATNet-C-FGSM, ATNet-C-BIM10UT and ATNet-C-BIM10RT.

These trojan networks were used to attack other three target networks, CNN-small, Alexnet and VGG9, which were trained on the vanilla CIFAR10 dataset. The results are shown in Table~\ref{tab:transfer}.

In Table~\ref{tab:transfer}, the trojan networks showed good transferability across different target networks. The predictions of the target networks were relatively hard to be influenced by such small perturbations when not being infected by the trojan networks. However, when infected by the trojan networks, all target networks were much easier to attack, especially when we used C-BIM10RT to craft the adversarial examples. For instance, the accuracy of the VGG9 was $83.95\%$ when it was attacked by BIM10RT, while the accuracy dropped to $7.70\%$ when it was infected by ATNet-BIM10RT and attacked by C-BIM10RT (10-iteration targeted attack method as BIM10RT).

\begin{table*}
  \scriptsize
\centering
\caption{the results of the target network Resnet50AT on Imagenet with the trojan network}
\begin{tabular*}{1.0\linewidth}{@{\extracolsep{\fill}}lllllll}
\hline
\multicolumn{2}{l}{\multirow{2}{*}{The target network}}  & Clean  & FGSM   & BIM10UT   & \multicolumn{2}{c}{BIM10RT} \\
\multicolumn{2}{l}{}                                     & acc/\% & acc/\% & acc/\%    & acc/\%       & suc/\%       \\ \hline
\multicolumn{2}{l}{Resnet50AT(No Trojan)}                                     & 59.06  & 37.88  & 35.92     & 56.68        & 0.36         \\ \hline
\multicolumn{2}{l}{\multirow{2}{*}{The trojan networks}} & Clean  & C-FGSM & C-BIM10UT & \multicolumn{2}{c}{C-BIM10RT} \\
\multicolumn{2}{l}{}                                     & acc/\% & acc/\% & acc/\%    & acc/\%       & suc/\%       \\ \hline
\multirow{2}{*}{ATNet-C-BIM10RT}           & on          & 58.96  & 3.26   & 0.30      & 0.52         & 93.16        \\
                                           & off         & 59.06  & 59.08  & 57.88     & 59.10        & 0.06         \\ \hline
\end{tabular*}
\label{tab:imagenet}
\end{table*}

\begin{table*}
\scriptsize
\centering
\caption{the results of the target network Resnet50AT on unseen ImageNet categories with the trojan network}
\begin{tabular*}{1.0\linewidth}{@{\extracolsep{\fill}}lllllll}
\hline
\multicolumn{2}{l}{\multirow{2}{*}{Trojan network}} & Clean  & C-FGSM  & C-BIM10UT & \multicolumn{2}{c}{C-BIM10RT} \\
\multicolumn{2}{l}{}               & acc/\% & acc/\% & acc/\%  & acc/\%        & suc/\%       \\ \hline
\multirow{2}{*}{ATNet-C-BIM10RT}             & on             & 58.48  & 3.28   & 0.56    & 0.88          & 91.68        \\
                                     & off            & 58.96  & 59.12  & 57.52   & 58.80         & 0.00         \\ \hline
\end{tabular*}
\label{tab:split}
\end{table*}

\begin{figure*}[b]
  \centering
  \subfigure[]{
  \includegraphics[width=0.32\linewidth]{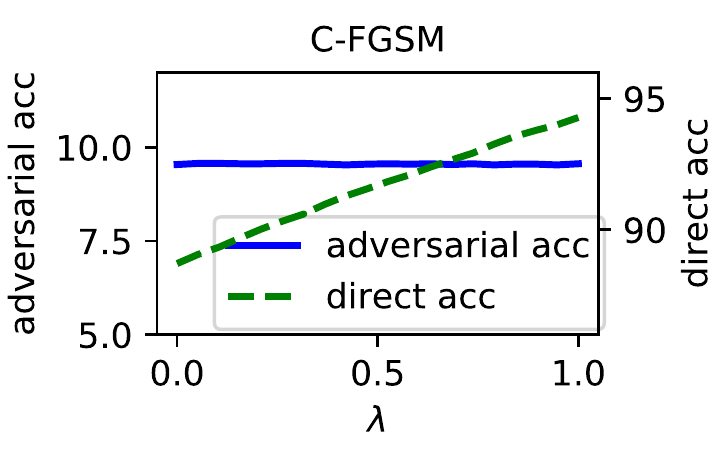}
  \label{fig:PFGSM}}
    \subfigure[]{
  \includegraphics[width=0.32\linewidth]{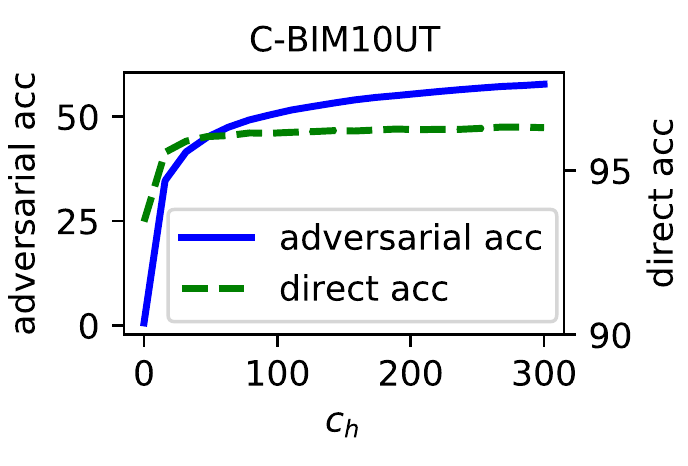}
  \label{fig:BIMUT}}
  \subfigure[]{
  \includegraphics[width=0.32\linewidth]{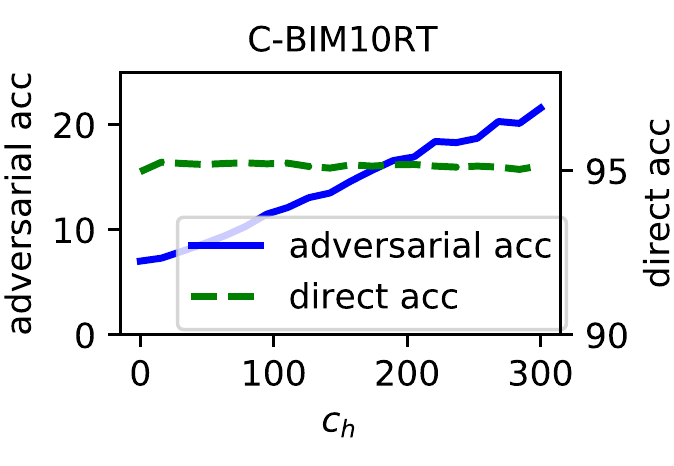}
  \label{fig:BIM}}
  \caption{Sensitivity results of the hyper-parameters in the C-FGSM and C-BIM10-K algorithms. \subref{fig:PFGSM} The performance of the C-FGSM with different $\lambda$. When $\lambda=0$, it is equivalent to FGSM. \subref{fig:BIMUT} The performance of the C-BIM10UT with different $c_h$. \subref{fig:BIM} The performance of the C-BIM10RT with different $c_h$.}
\end{figure*}

We had several other observations. First, for any target network paired with any trojan network, the adversarial examples crafted by the C-BIM10RT algorithm were more effective to attack the target network than those crafted by the C-FGSM and C-BIM10UT. For instance, The accuracies of VGG9 paired with ATNet-BIM10RT were $38.78\%$, $49.40\%$ and $7.70\%$, respectively. Therefore, we recommend using the C-BIM10RT to craft adversarial exmaples to attack the target networks. Second, when using C-BIM10RT to attack CNN-small and VGG9, the ATNet-C-BIM10RT was the strongest among the three trojan networks; but when using C-BIM10RT to attack Alexnet, the ATNet-C-FGSM was the strongest. The accuracies of CNN-small/Alexnet/VGG9 were $3.39\%/14.24\%/10.87\%$, $1.11\%/19.80\%/9.92\%$ and $0.89\%/28.83\%/7.70\%$ when applying ATNet-C-FGSM, ATNet-C-BIM10UT and ATNet-C-BIM10RT, respectively.


Finally, there was an unusual result that the attack ability of the 10-iteration attack method S-IM10UT seemed to be weaker than the one-step attack algorithm C-FGSM (see the accuracies in columns \emph{C-FGSM} and \emph{C-BIM10UT} of Table~\ref{tab:transfer}). We attributed this to the non-smoothness of the model’s loss surface, which will be discussed in Section \ref{sec:amplification}.


\subsection{Results on ImageNet}

We used Resnet50 as the target network on ImageNet. However, we found that the Resnet50 trained on the clean ImageNet images was too vulnerable as it could be broken easily by the adversarial examples crafted by standard algorithms. To demonstrate the amplification ability of the trojan network, we used the adversarial training method to train the target network. The performance of the adversarially trained Resnet50 on the clean examples and the adversarial examples are shown in Table~\ref{tab:imagenet}. The target network was robust to adversarial attacks without the trojan network. The success rate of the BIM10RT attack was only $0.36\%$.

We used C-BIM10RT to generate adversarial examples when training ATNet. The performance of the trojan network is shown in Table~\ref{tab:imagenet}. When the trojan network infected the target network, the success rate of the C-BIM10RT attack was $93.16\%$. It indicated that the trojan network can successfully break the target network.

In section~\ref{section:transfer} we have shown the generalization ability of the trojan network across different target networks on CIFAR10. Here we investigated the generalization ability of ATNet across different categories of examples on ImageNet. Specifically, we asked whether the trojan networks can influence the prediction accuracy of the target network on categories that have not been used for training the trojan networks.

We collected images from ImageNet as previously did. We divided the $1000$ categories into two parts: $750$ categories for training and the other $250$ categories for the test. We randomly chose five images for each category and constructed a training set with $3750$ images and a test set with $1250$ images. Note that the categories of training and test sets were not overlapped. We trained ATNet on the training set. The performance of the target network on the test set when infected by the trojan network is shown in Table~\ref{tab:split}. The results showed that ATAttack was able to attack the target network on the unseen categories.

\subsection{Parameter Sensitivity of the Concealable Adversarial Attack Methods}
\label{section:parameter}

In this section, we study the parameter sensitivity of the adversarial attack methods that craft the concealable adversarial examples. Note that there is a hyper-parameter $\lambda$ in C-FGSM (see Eq.~\eqref{eq:PFGSM}) and a hyper-parameter $c_h$ in C-BIM-K (see Eq.~(\ref{eq:BIM})). When $\lambda=0$, the C-FGSM is equivalent to the original FGSM. Similarly, when $c_h=0$, C-BIM-K is equivalent to the original BIM-K.

We used ATNet-C-FGSM, ATNet-C-BIM10UT and ATNet-C-BIM10RT as the trojan networks and used ResNet18 as the target network. All networks were fixed in this section, while the attack methods with different values of $\lambda$ and $c_h$ were evaluated.

\begin{figure*}[b]
  \centering
  \subfigure[]{
  \includegraphics[width=0.44\linewidth]{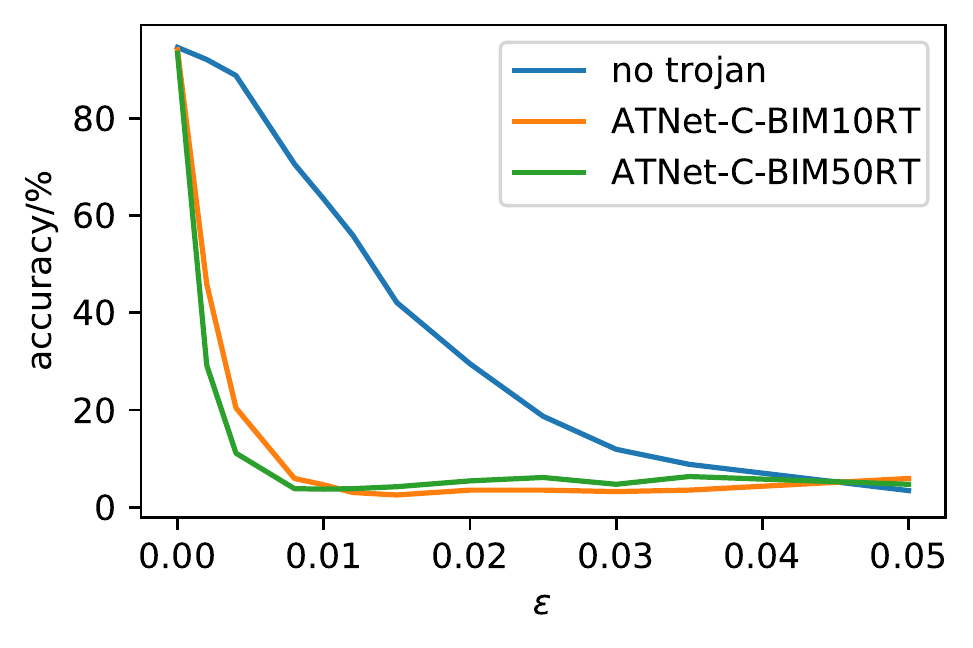}
  \label{fig:accBIM10}}
  \subfigure[]{
  \includegraphics[width=0.44\linewidth]{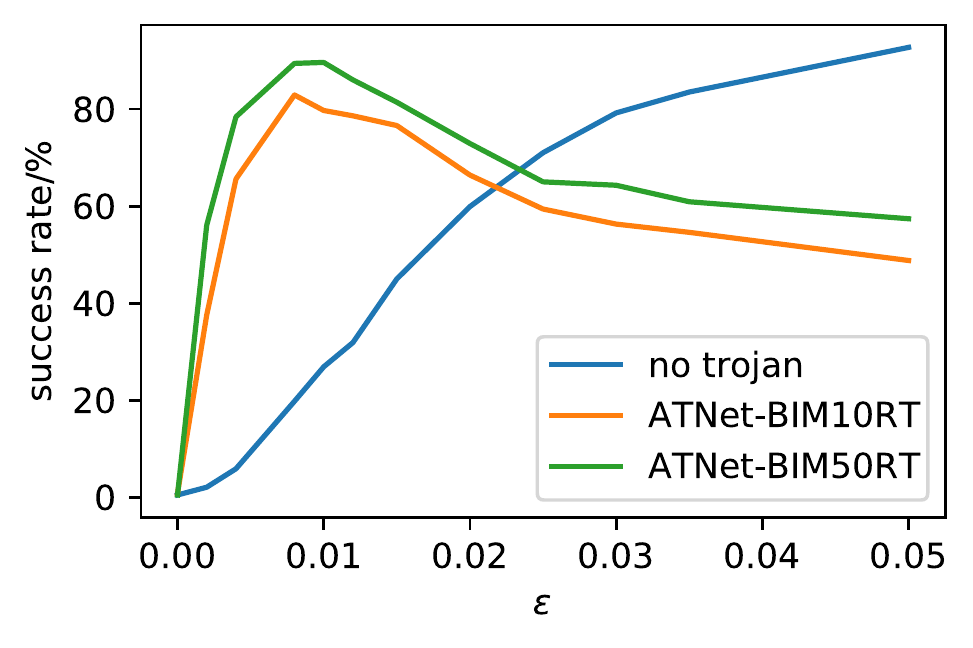}
  \label{fig:sucBIM10}}
  \subfigure[]{
  \includegraphics[width=0.44\linewidth]{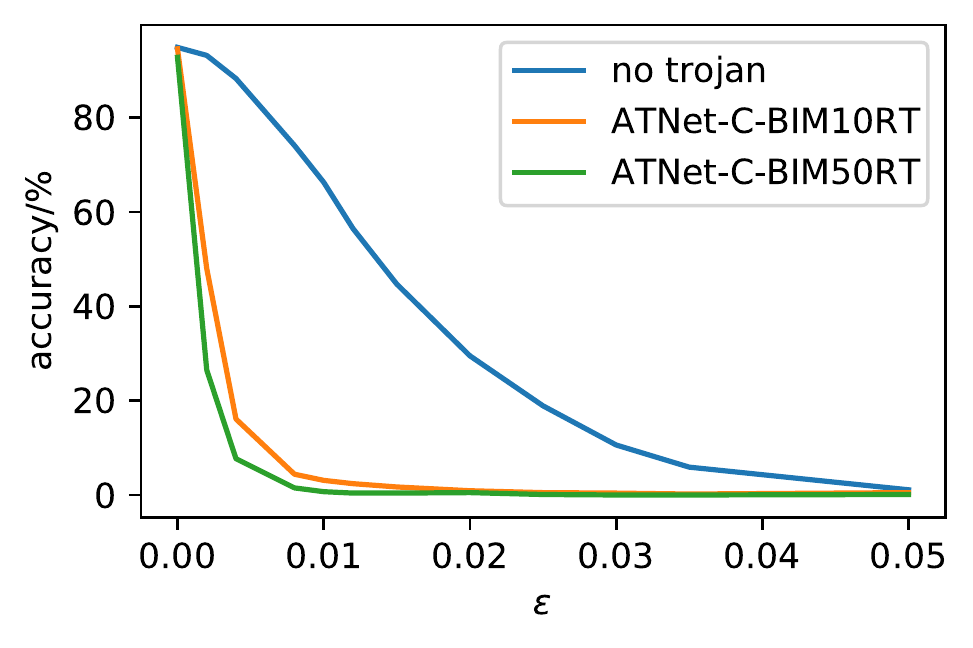}
  \label{fig:accBIM50}}
  \subfigure[]{
  \includegraphics[width=0.44\linewidth]{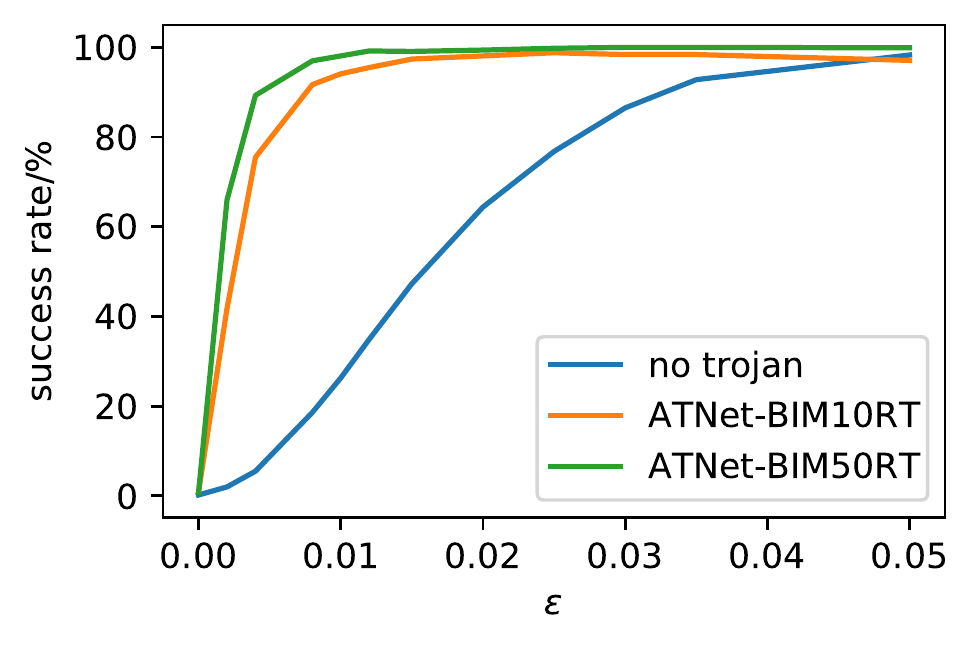}
  \label{fig:sucBIM50}}
  \caption{Visualization of the amplification ability. The horizontal axes are the upper bound $\epsilon$ of the perturbations. \subref{fig:accBIM10} The accuracy of Resnet18 on the examples crafted by C-BIM10RT. \subref{fig:sucBIM10} The success rate of the attack method C-BIM10RT. \subref{fig:accBIM50} The accuracy of Resnet18 on the examples crafted by C-BIM50RT. \subref{fig:sucBIM50} The success rate of the attack method C-BIM50RT.}
  \label{fig:eps_study}
\end{figure*}

We varied $\lambda$ when evaluating C-FGSM. See Fig. \ref{fig:PFGSM}. The \emph{adversarial accuracy} represents the accuracy of the target network on the adversarial examples when the trojan network is switched on. The \emph{direct accuracy} represents the accuracy when the trojan network is switched off. As shown in the figure, when $\lambda$ increased from $0$ to $1$, the direct accuracy increased while the adversarial accuracy changed little. C-FGSM with $\lambda=0$, which is equivalent to the original FGSM, had a low direct accuracy less than $90\%$, while C-FGSM with $\lambda=1$ had a direct accuracy close to the original clean accuracy, $95.00\%$ (Table \ref{table:CIFAR10_trojan}). It indicated that the original FGSM failed to meet the concealment requirement, while the C-FGSM with $\lambda=1$ could. Therefore we chose $\lambda=1$ in all the experiments reported in the other sections. It corresponds to the case that the adversarial perturbation is exactly along the direction of $-g_\perp$ according to (\ref{eq:PFGSM}).

When evaluating C-BIM-K, we increased $c_h$ from $0$ to $300$ and employed C-BIM10UT and C-BIM10RT. See Fig.~\ref{fig:BIMUT}. When we employed C-BIM10UT, the direct accuracy increased a little when $c_h$ increased. However, the adversarial accuracy increased quickly in the meantime. See Fig.~\ref{fig:BIM}. When we employed C-BIM10RT, the direct accuracy only slightly fluctuated when $c_h$ increased, while the adversarial accuracy increased quickly. Note that the lower adversarial accuracy indicates a stronger attack. Concluded from Fig.~\ref{fig:BIMUT} and Fig.~\ref{fig:BIM}, the C-BIM-K with $c_h>0$ was helpful to satisfy the concealment requirement but could bring negative effects to the attack requirement. Without loss of generality, we chose $c_h=0$ when employing C-BIM10UT and C-BIM10RT in all the experiments reported in the other sections.

\subsection{Analysis of the Amplification Ability}
\label{sec:amplification}

In this section, we show how the trojan networks amplified the weakness of the target networks. We used Resnet18 as the target network and used ATNet-C-BIM10RT as the trojan network. We used C-BIM10RT to craft adversarial examples. The results are obtained by evaluating the target network on CIFAR10. See the blue and the orange curves in Fig.~\ref{fig:accBIM10} and Fig.~\ref{fig:sucBIM10}. When the trojan network was switched on, the accuracy of the target network decreased much faster as the perturbation increased. These results demonstrated the amplification ability of ATNet.

Unexpectedly, we found that the success rate in Fig.~\ref{fig:sucBIM10} decreased as the perturbation $\epsilon$ in Eq.~(\ref{eq:imperceptibility}) increased. However, a perturbation that satisfies a smaller constrain also satisfies a larger one. Thus, the success rate should increase when $\epsilon$ increased. We made an assumption that C-BIM10RT has encountered inadequate optimization, i.e., the algorithm C-BIM10RT with 10 iterations to satisfy the requirements might not be able to find strong enough adversarial examples. Therefore, we used $K=50$ in C-BIM-K (denoted as C-BIM50RT) to attack the target network which is still infected by ATNet-C-BIM10RT. See the orange curve in Fig.~\ref{fig:sucBIM50}. The success rate indeed monotonously increase when $\epsilon$ increased.

Since C-BIM10RT was found insufficient to craft strong enough adversarial examples, there came up a further question: is C-BIM10RT enough to generate the adversarial examples when we train the trojan network? We trained the ATNet to attack Resnet18 with C-BIM50RT to generate adversarial examples (the obtained trojan network was named by ATNet-C-BIM50RT). See green curves in Fig. \ref{fig:eps_study}. We found that ATNet-C-BIM50RT only performed slightly better than ATNet-C-BIM10RT, though the training time of the former was roughly five times the time of the latter. It indicated that C-BIM10RT was enough to train a strong trojan network. Thus we recommend training the trojan network with C-BIM10RT rather than using C-BIM-K with more iteration steps.

\begin{figure*}[t]
  \centering
  \subfigure[]{
  \includegraphics[width=0.44\linewidth]{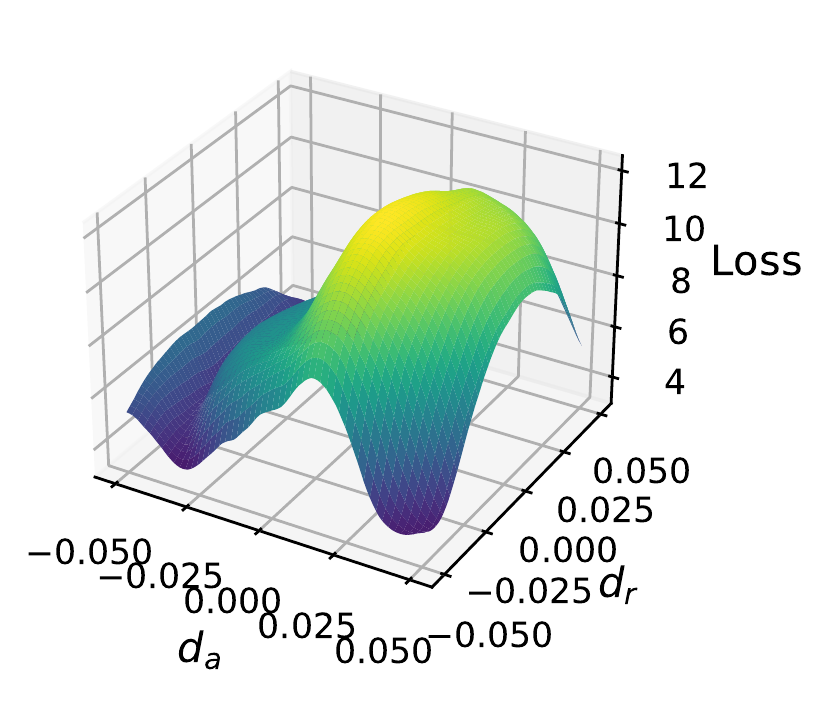}
  \includegraphics[width=0.44\linewidth]{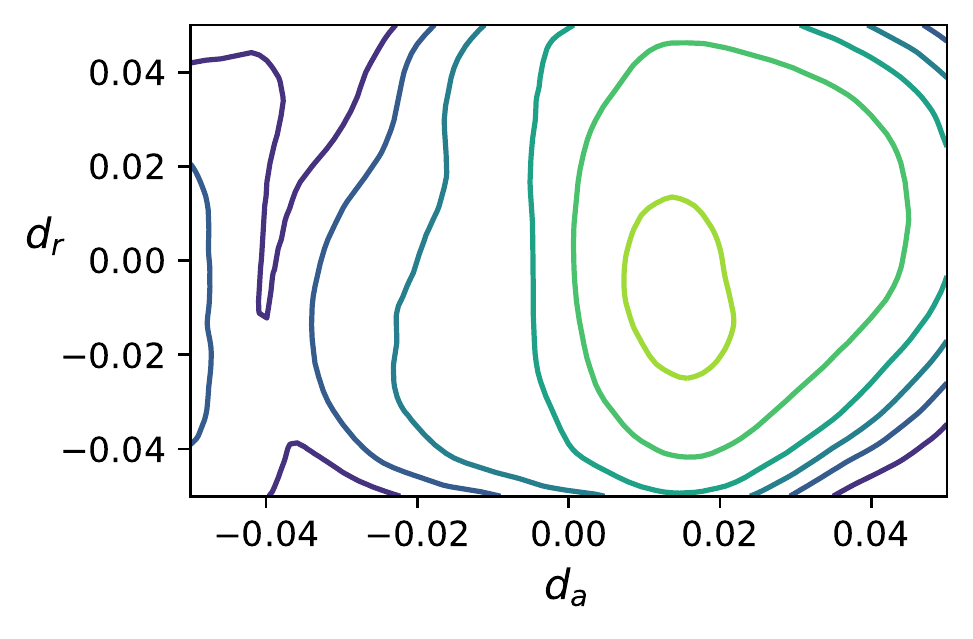}
  \label{fig:losssurf_off}}
  \subfigure[]{
  \includegraphics[width=0.44\linewidth]{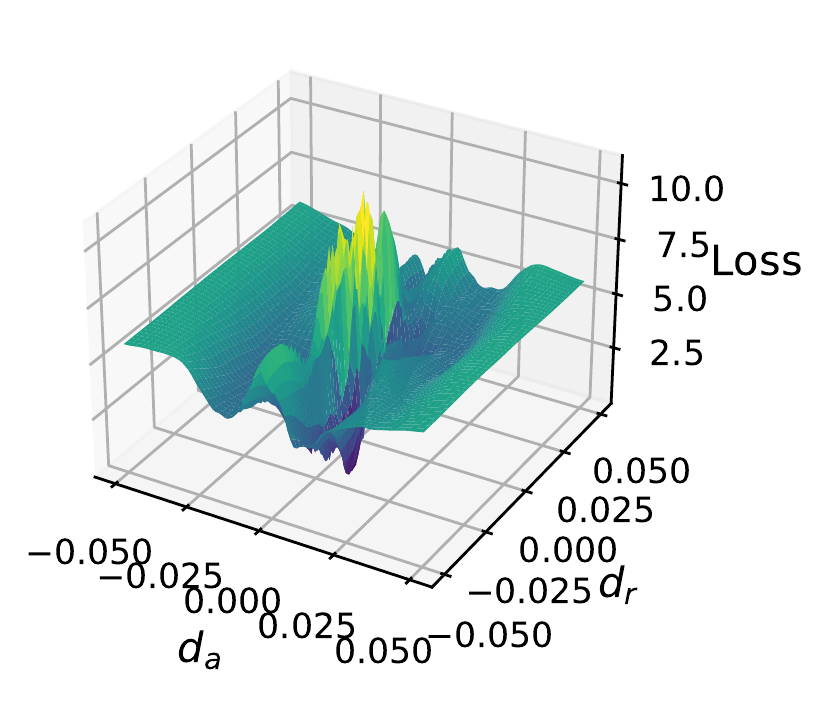}
  \includegraphics[width=0.44\linewidth]{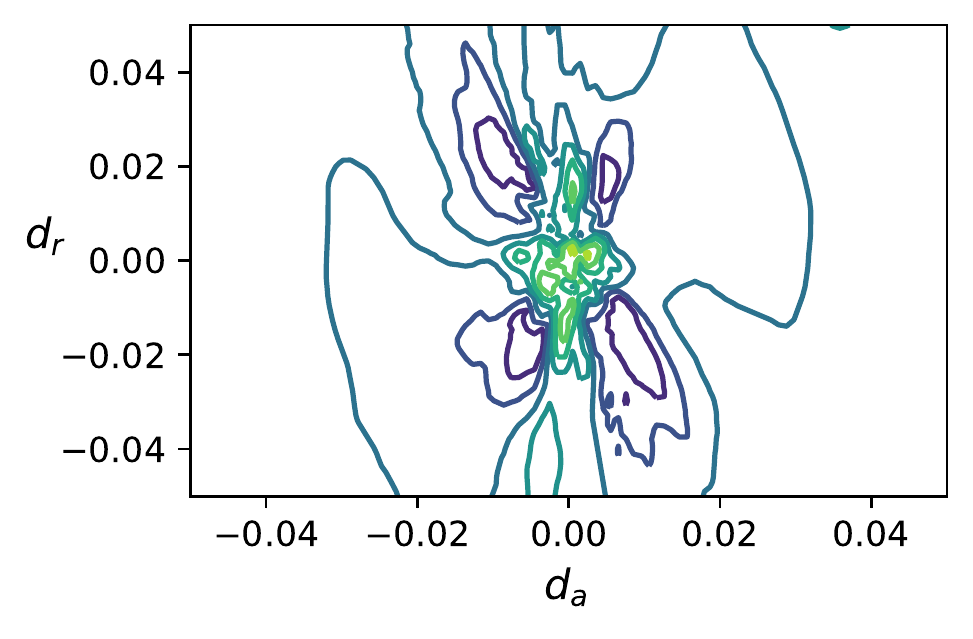}
  \label{fig:losssurf_on}}

  \caption{The loss surface of the target network when the trojan network is switched on or off. \subref{fig:losssurf_off} The trojan network is switched off. \subref{fig:losssurf_on} The trojan network is switched on. The left of \subref{fig:losssurf_off} and \subref{fig:losssurf_on} are the surface maps of the loss function when the trojan network is switched on or off. The right of \subref{fig:losssurf_off} and \subref{fig:losssurf_on} are the corresponding contour maps.}
  \label{fig:loss_surface}
\end{figure*}

We plot the loss surface with respect to two axes in Fig.~\ref{fig:loss_surface}. One axis $d_a$ is along the direction of the sign of the gradient of the loss function, $\mathrm{Sign}(\nabla(\mathrm{CEloss}(F(x), l)))$. The other axis $d_r$ is along the sign of a random direction, $\mathrm{Sign}(\mathrm{Normal_n})$, where $\mathrm{Normal_n}$ indicates an $n$-dimensional random vector sampled from a normal distribution. The mechanism of gradient-based adversarial attack algorithms like C-BIM-K can be regarded as searching for a local minimum point on the surface of the loss function. When the trojan network was switched on, the loss surface was much more rugged. It can give some explanations for some of the above observations. First, the loss changes rapidly around the original point when the trojan network is switched on. Therefore, the adversarial attack algorithms are more likely to find valid adversarial examples when the perturbation is small, i.e., the target network became more vulnerable under adversarial attacks. Second, the gradient also changes rapidly around the original point when the trojan network is switched on. Since C-BIM-K takes $K$ steps with a certain size along the direction of gradients, it may be hard to find a point whose loss is small enough (the point corresponds to a strong enough adversarial example) when the step size is too large. Therefore, the success rate of C-BIM10RT can decrease when $\epsilon$ increases (the algorithm takes a larger step when $\epsilon$ is larger). C-BIM50RT, which takes more steps and smaller step size than C-BIM10RT, can still find a strong enough adversarial example when $\epsilon$ is large.

\section{Conclusion}

We have introduced a new attack method for DNNs named ATAttack. A trojan network ATNet is placed before the target DNN to transform the input, such that the altered input can mislead the target DNN. ATNet performs the attack by amplifying the inherent weakness of the target network. The target network will be much more vulnerable to adversarial examples when being infected by the trojan network.

Experiments showed that ATNet was effective in attacking typical DNNs. In addition, ATNet trained with one target network was able to mislead other target networks. It suggests that there could be something in common with the inherent weakness of different target networks.

Since ATAttack does not change the structure or parameters of the target DNN, the trojan network can be implanted in a program that is independent of the target network. This attack method poses a new threat to DNNs. In developing safe and robust DNNs, this threat should be taken into account.








\section*{Acknowledgements}

This work was supported in part by the National Natural Science Foundation of China under Grant U19B2034 and Grant 62061136001.

\bibliography{refs}

\end{document}